\documentclass[aps,prl,twocolumn,floats]{revtex4-2} 
\usepackage{amsmath}
\usepackage{graphicx}
\usepackage{dcolumn}
\usepackage{bm}
\usepackage{amssymb}
\usepackage{latexsym}
\usepackage{color}
\usepackage{subfigure}
\usepackage{ulem}

\begin{document}

\title{Atom beam-splitter with internal state selection using spin-dependent optical standing wave potentials}

\author{Igor Kuzmenko$^{1,2}$, Yshai Avishai$^{1,3}$, Y. B. Band$^{1,2,4}$}

\affiliation{
  $^1$Department of Physics,
  Ben-Gurion University of the Negev,
  Beer-Sheva 84105, Israel
  \\
  $^2$Department of Chemistry,
  Ben-Gurion University of the Negev,
  Beer-Sheva 84105, Israel
  \\
  $^3$Yukawa Institute of Theoretical Physics, Kyoto, Japan\\
  $^4$The Ilse Katz Center for Nano-Science,
  Ben-Gurion University of the Negev,
  Beer-Sheva 84105, Israel
  }

\begin{abstract}
We propose an atom beam splitter that enables the manipulation of the internal spin state of the atoms in the output beams using a spin-dependent optical potential. The utility of such an atom beam splitter is demonstrated through its application in measuring the Aharonov-Casher phase of atoms subjected to a constant homogeneous electric field, thereby enabling measurement of the electric field strength.
\end{abstract}

\maketitle

{\it Introduction}---Atom beam splitters (BSs) are crucial components in atom quantum optics and atomic physics in general, enabling the manipulation and study of atoms at the quantum level. They are essential for various applications, including atom interferometry, quantum computing, quantum simulation, and the manipulation of matter waves, which in turn have a multitude of applications in physics \cite{Cronin_09}.  One method of producing an atom (or molecule \cite{Brand_20}) BS is to use a standing wave light field. The Bragg scattering from the standing wave light field can result in a momentum change of atoms in an atomic beam \cite{Martin_88, Kozuma_99}, thereby effectively splitting the beam of atoms into two different beams. Atom interferometers are utilized for the purpose of testing fundamental physics principles, as well as for the measurement of physical constants and inertial forces \cite{Cronin_09, atom_inferf_14, Parker_18, Morel_20}.  Moreover, atom interferometers are utilized as high-accuracy accelerometers, gravimeters \cite{Peters_01, Geiger_20, Kitching_11, Dickerson-2013}, gravity gradiometers (instruments for measuring gravity gradients) \cite{Kitching_11}, gyroscopes \cite{Geiger_20, Kitching_11, Dickerson-2013} and magnetometers \cite{Ockeloen_13}.  They have also been used to test the equivalence principle, which stipulates the equivalence of gravitational and inertial mass \cite{Asenbaum_20}.  Using large-momentum-transfer atom interferometers with elastic Bragg scattering of light waves reduces the systematic error of interferometers \cite{Kirsten_23, Peng-25}.

\begin{figure}
\centering
\includegraphics[width=0.8\linewidth,angle=0] {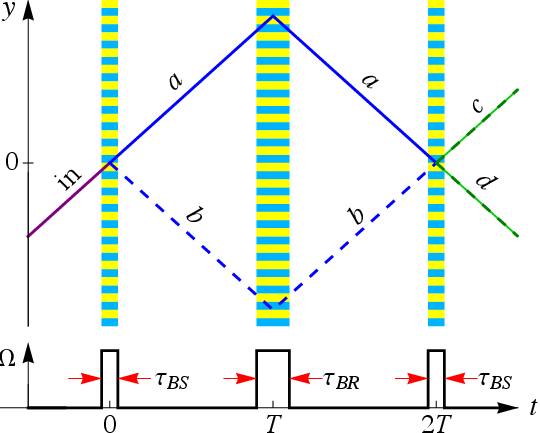}
\caption{\footnotesize
Spin-dependent Mach-Zender atom interferometer that employs two SDO BSs. Top frame: The incoming atomic beam (``in'', solid purple line) is split by the first SDO BS centered at time $t = 0$ into beams $a$ and $b$ (solid and dashed blue lines, respectively).  The beam reflector (BR) centered at $t = T$ reflects beams $a$ and $b$, by changing the sign of the $y$ component of the beam momentum.  Another SDO BS centered at $t = 2 T$ splits each of the beams $a$ and $b$ into beams $c$ and $d$ (green lines) such that each of the beams $c$ and $d$ is a linear combination of atoms originating from beams $a$ and $b$.  For the BSs, yellow and turquoise indicate the area where $B(y)$ [see Eq.~(\ref{eq:H_v})] is positive and negative, respectively.  For the BR, yellow and turquoise indicate the area where $V(y)$ [see Eq.~(\ref{eq:H_s})] is positive and negative.  Bottom frame: Rabi frequency $\Omega$ versus time.  The duration of the BS pulses with light frequency $\omega_0$ is $\tau_{\rm BS}$, and the duration of the BR pulse with light frequency $\omega_1$ is $\tau_{\rm BR}$.  The corresponding Rabi frequencies, $\Omega_{\rm BS}$ for the BS, and $\Omega_{\rm BR}$ for the BR are taken to be equal.}
\label{Fig:Bragg-splitter}
\end{figure}

Here we propose a new type of atom BS that employs a spin-dependent optical (SDO) potential \cite{SOI-EuroPhysJ-13} formed by standing wave light beams whose polarizations are arranged in a manner that enables the manipulation of the internal spin state of the atoms in the output beams.  Atom BSs of this nature have potential applications in a variety of scientific instruments, including atom interferometers with path-entangled states, optical Ramsey-Bord\'e interferometers, and Raman interferometers \cite{Cronin_09}.  A Mach-Zender atom interferometer, as depicted in Fig.~\ref{Fig:Bragg-splitter}, serves as a paradigm for the application of our novel atom BS. This apparatus facilitates the measurement of the Aharonov-Casher phase of atoms subjected to a static homogeneous electric field and therefore the measurement of the electric field strength \cite{AC_84, Kuzmenko_25, Kuzmenko_25b}.  This application demonstrates the necessity of using a SDO BS.

SDO lattice potentials for optical traps are described in Ref.~\cite{SOI-EuroPhysJ-13}, and have been used to coherently manipulate atomic spins \cite{Yang_2017}, and to study topological properties of matter \cite{Kuzmenko_24}.  Atoms deeply trapped in a SDO potential behave as elementary quantum rotors \cite{Kuzmenko_19, Szulim_22}.  Here we show that SDO potentials can be used to make atom BSs which change the internal state of atoms.

A Ramsey-Bord\'e interferometer \cite{Cronin_09, Borde_89} is a specific type of atom interferometer which puts the atoms into a superposition of ground and excited internal states, and is utilized for high-precision measurements, particularly in laser frequency stabilization and tests of fundamental physics.  Our BS and interferometer also puts the atoms into a superposition of internal spin states using SDO lattices, whereas the Ramsey-Bord\'e interferometer does not affect the internal spin state of the atoms.

{\it Dynamical polarizability of atoms}---For simplicity, we consider atoms with a ground state of $J = 1/2$, e.g., alkali atoms.  These atoms possess only scalar and vector polarizabilities, and vanishing tensor polarizability \cite{SOI-EuroPhysJ-13, comment-tensor}.  The ground state of alkali atoms is the ${}^{2}S_{1/2}$ state with the azimuthal quantum number $L = 0$, the spin quantum number $S = \frac{1}{2}$, and the total electronic angular momentum quantum number $J = \frac{1}{2}$.  The hyperfine interaction splits the ground state into states with total angular momentum of the atom $F = I \pm \frac{1}{2}$, where $I$ is the nuclear spin quantum number.  The first excited state is a ${}^{2}P$ state with $L = 1$.  Spin-orbit interaction splits the first excited state into a ${}^{2}P_{1/2}$ state with $J = \frac{1}{2}$, and a ${}^{2}P_{3/2}$ state with $J = \frac{3}{2}$.
Consider an alkali ground state atom illuminated by light with frequency $\omega$ far detuned from the resonance frequencies $\omega_{D_1}$ and $\omega_{D_2}$ of the ${}^{2}S_{1/2} \to {}^{2}P_{1/2}$ and ${}^{2}S_{1/2} \to {}^{2}P_{3/2}$ transitions, respectively.  (See the End Matter for details regarding the scalar polarizability $\alpha_s (\omega)$ and the vector polarizability $\alpha_v (\omega)$.)

\begin{figure}
\centering
  \includegraphics[width=0.8\linewidth,angle=0] {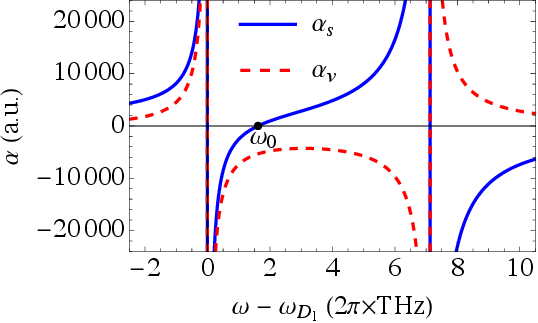}
\caption{\footnotesize
The scalar and vector polarizabilities of ${}^{87}$Rb atoms in the ground state with $F = 1$.}
\label{Fig:Rb-F1-polarizability}
\end{figure}

Figure~\ref{Fig:Rb-F1-polarizability} shows the scalar polarizability $\alpha_s (\omega)$ and the vector polarizability $\alpha_v (\omega)$ of ${}^{87}$Rb in the ground state with $F = 1$ (the ground and first excited states of ${}^{87}$Rb atoms, as well as the $D_1$ and $D_2$ spectral lines  are described in Ref.~\cite{Steck-87Rb}).  For the frequency in the interval $\omega_{D_1} < \omega < \omega_{D_2}$, the vector polarizability is negative (see dashed red curve).  $\alpha_s (\omega) < 0$ for $\omega_{D_1} < \omega < \omega_0$, $\alpha_s (\omega) > 0$ for $\omega_0 < \omega < \omega_{D_2}$, and $\omega_0$ is the angular frequency at which $\alpha_s (\omega_0) = 0$ [see Fig.~\ref{Fig:Rb-F1-polarizability}], $\omega_0 = \omega_{D_1} + 2 \pi \times 1.61842$ THz.  The scalar and vector polarizabilities for the other alkali atoms are qualitatively similar to those shown in Fig.~\ref{Fig:Rb-F1-polarizability}.  For $\omega_{D_1} < \omega < \omega_{D_2}$, $\alpha_v (\omega) < 0$ for $F = I - 1/2$, and $\alpha_v (\omega) > 0$ for $F = I + 1/2$.   Moreover, there exists a frequency $\omega_0$ within the interval $(\omega_{D_1}, \omega_{D_2})$ such that $\alpha_s (\omega) = 0$ at $\omega = \omega_0$.

{\it Bragg diffraction by a SDO lattice}---Consider atoms illuminated by two counter-propagating laser pulses with the same frequency, $\omega$, and wave vectors ${\bf K}_u$ and ${\bf K}_d$, so that the electromagnetic field forms a standing wave pattern shown in Fig.~\ref{Fig:Bragg-splitter}.   The electric field of the pulses is $\boldsymbol{\mathcal E} ({\bf r}, t) = \frac{1}{2} \, {\bf E} ({\bf r}) \, e^{- i \omega t} + {\rm c.c.}$, where
\begin{equation}    \label{eq:electric-field}
  {\bf E} ({\bf r}) =
  E_0 \, \boldsymbol\xi_u \, e^{i {\bf K}_u \cdot {\bf r}} +
  E_0 \, \boldsymbol\xi_d \, e^{i {\bf K}_d \cdot {\bf r}} .
\end{equation}
Here $\boldsymbol\xi_u$ and $\boldsymbol\xi_d$ are the polarization unit vectors, and the electric field amplitude $E_0$ is real.

The optical lattice potential for a $J = 1/2$ atom illuminated by the light is given by the Stark Hamiltonian~\cite{SOI-EuroPhysJ-13}
\begin{eqnarray}   \label{eq:H_Stark}
  H_{\rm Stark} &=&
  - \frac{\alpha_s (\omega)}{4} \, {\bf E}^{*} ({\bf r}) \cdot {\bf E} ({\bf r})
    \nonumber \\ && -
    \frac{i \alpha_v(\omega)}{8 F} \, \big[ {\bf E}^{*} ({\bf r}) \times {\bf E} ({\bf r}) \big] \cdot {\bf F} ,
\end{eqnarray}
where $\alpha_s (\omega)$ and $\alpha_v (\omega)$ are the scalar and vector polarizabilities given in Eqs.~(\ref{eq:alpha_s}) and (\ref{eq:alpha_v}) in End Matter.  The first term in Eq.~(\ref{eq:H_Stark}) is the scalar potential experienced by the alkali atoms in the optical lattice, and the second term is the vector potential, sometimes referred to as the {\it fictitious magnetic field} potential \cite{CT-72}, ${\bf B}_f \equiv \frac{i \alpha_v(\omega)}{2 F} \, \big[ {\bf E}^{*} ({\bf r}) \times {\bf E} ({\bf r}) \big]$, because the coupling to ${\bf F}$ is just like the coupling to a real magnetic field.

{\it Spin-dependent optical lattice BS}---For constructing an atom BS using a $\pi/2$ pulse, one can take the laser angular frequency to be $\omega_0$ shown in Fig.~\ref{Fig:Rb-F1-polarizability}, so the scalar polarizability vanishes, leaving just the vector polarizability.  Consequently, the optical lattice potential is proportional to ${\bf F} \cdot {\bf B}_f$, which gives rise to spin flips. The wave vectors are ${\bf K}_u = K_0 \, \hat{\bf y}$ and ${\bf K}_d = - K_0 \ \hat{\bf y}$, where $K_0 = \omega_0 / c$, and the polarization unit vectors are $\boldsymbol\xi_u = \hat{\bf z}$ and $\boldsymbol\xi_d = \hat{\bf x}$.  The vector polarizability of ${}^{87}$Rb is $\alpha_v (\omega_0) = - 5339.29 ~ a_B^3$, where $a_B$ is the Bohr radius, and the SDO potential in Eq.~(\ref{eq:H_Stark}) takes the form,
\begin{equation}   \label{eq:H_v}
  H_{\rm Stark, BS} = \hbar \, \Omega_{\rm BS} \, \sin (2 K_0 y) \, F_y ,
\end{equation}
where the Rabi frequency $\Omega_{\rm BS}$ is
\begin{equation}   \label{eq:Rabi-frequency_v}
  \Omega_{\rm BS} = \frac{1}{\hbar} \, | \alpha_v (\omega_0)| E_0^2 .
\end{equation}
The  fictitious magnetic field is ${\bf B}_f \equiv 2 \hbar \, \Omega_{\rm BS} \, \sin (2 K_0 y) \, {\hat{{\bf y}}}$ \cite{CT-72}.

The total time-dependent Hamiltonian of the atoms is
\begin{equation}   \label{eq:H0+H_BS}
  H (t) = H_0 + H_{\rm Stark, BS} \, \Theta \Big( \frac{\tau_{\rm BR}}{2} - |t| \Big) ,
\end{equation}
where $H_0 = - \frac{\hbar^2}{2 M} \, \nabla^2$ and $\Theta (\bullet)$ is the Heaviside step function.  The atomic time-dependent wave function $\Psi ({\bf r}, t)$ is found from the Schr\"odinger equation:
\begin{equation}   \label{eq:Schrodinger-BS}
  i \hbar \frac{\partial \Psi ({\bf r}, t)}{\partial t} = H(t) \Psi ({\bf r}, t) .
\end{equation}
We take the initial condition of the atomic wave function at time $t = - \tau_{\rm BS}/2$ to be $\Psi ({\bf r}, - \tau_{\rm BS}/2) = \chi_1 \, \psi_{{\bf k}_0} ({\bf r})$, where the basis spin wave functions $\chi_{m}$ with $m = 0, \pm 1$ are eigenfunctions of $F_z$ with eigenvalues $m$.  The spatial wave function is $\psi_{{\bf k}_0} ({\bf r}) = e^{i {\bf k}_0 \cdot {\bf r}}$, where the initial wave vector is ${\bf k}_0 = (k_x, K_0, 0)$.  The $y$ component of ${\bf k}_0$ must equal $K_0$ (or an integer multiple of $K_0$) in order to satisfy the Bragg scattering condition.  The wave functions of the states resulting from Bragg scattering are $\chi_{m} \psi_{{\bf k}_n} ({\bf r})$, where $\psi_{{\bf k}_n} ({\bf r}) = e^{i {\bf k}_n \cdot {\bf r}}$, with ${\bf k}_n = (k_x, (2 n + 1) K_0, 0)$, and integer $n$.  The wave function $\psi_{{\bf k}_n} ({\bf r})$ is an eigenfunction of $H_0$, and the corresponding energy is
$\epsilon_{k_n} = \frac{\hbar^2 | {\bf k}_n |^2}{2 M} = \frac{\hbar^2 k_0^2}{2 M} + 4 \, {\mathcal E}_{0} \, n \, (n + 1)$, where $k_n = | {\bf k}_n |$, and the recoil energy is ${\mathcal E}_{0} = \tfrac{\hbar^2 K_0^2}{2 M}$.  For $n = -1$, ${\bf k}_{- 1} = (k_x, -K_0, 0)$, and $\epsilon_{k_{-1}} = \epsilon_{k_0}$.  Furthermore, for $n \geq 1$, $\epsilon_{k_n} = \epsilon_{k_{- n - 1}}$, and the energy difference $\epsilon_{k_n} - \epsilon_{k_0} = 4 \, {\mathcal E}_{0} \, n \, (n + 1) \geq 8 {\mathcal E}_{0}$.

The matrix elements of $H_{\rm Stark}$ are
\begin{eqnarray}
  {\mathcal B}_{{\bf k}_n m, {\bf k}_{n'} m'} &\equiv&
  \int \psi_{{\bf k}_n}^{*} ({\bf r}) \chi_{m}^{\dag} \, H_{\rm Stark} \, \chi_{m'} \psi _{{\bf k}_{n'}} ({\bf r}) \, d^3 {\bf r}
  \nonumber \\ &=&
  - \frac{i \, \hbar \, \Omega_{\rm BS}}{2} \, F^{y}_{m, m'} \, \big[ \delta_{n, n'+1} - \delta_{n, n'-1} \big] , 
\end{eqnarray}
where $F^{y}_{m, m'} \equiv \chi_{m}^{\dag} F_y \chi_{m'} = - \frac{i}{\sqrt{2}} \, ( \delta_{m, m'-1} - \delta_{m, m'+1} )$.

As shown in the End Matter, the Schr\"odinger equation (\ref{eq:Schrodinger-BS}) reduces to the following when $\hbar \Omega_{\rm BS} \ll {\mathcal E}_{0}$,
\begin{equation}   \label{eq:for-X-BS-2-levels}
  i \,
  \frac{d}{d t}
  \left( \! \! \!
    \begin{array}{c}
      {\mathcal X}_{0} (t) \\ {\mathcal X}_{-1} (t)
    \end{array}
  \! \! \!
  \right) = \frac{i}{2}
  \left( \! \!
    \begin{array}{cc}
      0 & \Omega_{\rm BS} F_y
      \\
      - \Omega_{\rm BS} F_y & 0
    \end{array}
  \! \!
  \right) \,
  \left( \! \! \!
    \begin{array}{c}
      {\mathcal X}_{0} (t) \\ {\mathcal X}_{-1} (t)
    \! \! \!
    \end{array}
  \right) ,
\end{equation}
where ${\mathcal X}_{n} (t)$ and its initial condition at $t = - \frac{\tau_{\rm BS}}{2}$ are given in the End Matter.
%
The solution of Eqs.~(\ref{eq:for-X-BS-2-levels}) satisfying  the initial conditions is ${\mathcal X}_{0} (t) = \cos^2( \tfrac{\Omega_{\rm BS} {\tilde t}}{4}) \chi_1 + \sin^2( \tfrac{\Omega_{\rm BS} {\tilde t}}{4}) \chi_{-1}$,
${\mathcal X}_{-1} (t) = - i \, \sin( \frac{\Omega_{\rm BS} {\tilde t}}{2} ) \chi_0$, where ${\tilde t} = t + \tau_{\rm BS}/2$.  The duration of the BS pulse is $\tau_{\rm BS} = \frac{\pi}{2 \Omega_{\rm BS}}$, and 
${\mathcal X}_{0} ( \frac{\tau_{\rm BS}}{2} )$ and ${\mathcal X}_{- 1} ( \frac{\tau_{\rm BS}}{2} )$ at $t = \frac{\tau_{\rm BS}}{2}$ are
\begin{equation}   \label{spin-wave-function-BS1}
  \left(
    \begin{array}{c}
      {\mathcal X}_{0} (\frac{\tau_{\rm BS}}{2} ) \\ {\mathcal X}_{- 1} ( \frac{\tau_{\rm BS}}{2} )
    \end{array}
  \right) =
  \left(
    \begin{array}{cc}
      {\mathcal V}_{T} & {\mathcal V}_{R}
      \\
      - {\mathcal V}_{R} & {\mathcal V}_{T}
    \end{array}
  \right) \,
  \left(
    \begin{array}{c}
      \chi_1 \\ 0
    \end{array}
  \right) ,
\end{equation}
where
\begin{equation}   \label{eq:U_BS-V_BS}
  {\mathcal V}_{T} =
  \frac{\sqrt{2} + 1}{3} \, {\mathbb I} - \frac{\sqrt{2} - 1}{2 \sqrt{2}} \, Q_{y, y} ,
  \quad
  {\mathcal V}_{R} = \frac{1}{\sqrt{2}} \, F_y .
\end{equation}
Here ${\mathbb I}$ is the 3$\times$3 identity matrix, and the matrices $Q_{j, j'}$ are given by
\begin{equation}   \label{eq:quadrupole-def}
  Q_{j, j'} = F_j F_{j'} + F_{j'} F_j - \frac{2}{3} \, F (F+1) \, \delta_{j, j'} ,
\end{equation}
where $j$ and $j'$ are Cartesian indices and $F = 1$.  Note that ${\mathcal V}_{T}$ and ${\mathcal V}_{R}$ are 3$\times$3 hermitian matrices which satisfy the following properties:
$[{\mathcal V}_{T}, {\mathcal V}_{R}] = 0$ and ${\mathcal V}_{T}^{2} + {\mathcal V}_{R}^{2} = {\mathbb I}$.  Thus the 6$\times$6 matrix in the right-hand side of Eq.~(\ref{spin-wave-function-BS1}) is unitary.

After the pulse, the beam having spin wave function $\chi_{1}$ and wave vector ${\bf k}_0 = (k_x, K_0, 0)$ splits into two beams; one beam has spin wave function ${\mathcal X}_{0} (\frac{\tau_{\rm BS}}{2})$ and wave vector ${\bf k}_0$, and the other beam has spin wave function ${\mathcal X}_{-1} (\frac{\tau_{\rm BS}}{2})$ and wave vector ${\bf k}_{-1} = (k_x, -K_0, 0)$.

{\it Spin-dependent Mach-Zehnder interferometer}---We now construct a Mach-Zehnder interferometer using SDO BS pulses, as described above, and a spin-independent beam reflector (BR) pulse, as illustrated in Fig.~\ref{Fig:Bragg-splitter}.  
The subsequent sections will provide a detailed description of the BR.

{\it Spin-independent Bragg beam-reflector}---For constructing the BR pulse, one can take the laser frequency $\omega_1$ such that $\omega_0 < \omega_1 < \omega_{D_2}$, and far-detuned from the resonance frequency $\omega_{D_2}$.  The pulse wave vectors are ${\bf K}_{u} = K_1 \, \sin \theta_1 \, \hat{\bf x} + K_1 \, \cos \theta_1 \, \hat{\bf y}$ and ${\bf K}_{d} = K_1 \, \sin \theta_1 \, \hat{\bf x} - K_1 \, \cos \theta_1 \, \hat{\bf y}$, and the polarization unit vectors $\boldsymbol\xi_u = \boldsymbol\xi_d = \hat{\bf z}$, see Eq.~(\ref{eq:electric-field}).  Here $K_1 = \frac{\omega_1}{c}$, $\theta_1 = \arccos ( \frac{K_0}{K_1})$.  For example, the SDO lattice can be constructed with frequency $\omega_1 = \omega_{D_2} - 2 \pi \times 2.92011$~THz \cite{comment:omega_1}, for which $\theta_1 = 0.116496$~rad.  The optical potential in Eq.~(\ref{eq:H_Stark}) takes the form,
\begin{equation}   \label{eq:H_s}
  H_{\rm Stark, BR} = \hbar \Omega_{\rm BR} \big[ 1 + \cos (2 K_0 y) \big] ,
\end{equation}
where the Rabi frequency $\Omega_{\rm BR}$ is
\begin{equation}   \label{eq:Rabi-frequency_s}
  \Omega_{\rm BR} = \frac{1}{\hbar} \, | \alpha_s (\omega_1)| E_0^2 .
\end{equation}
The optical potential is switched on at $t = T_{-}$, and switched off at $t = T_{+}$, where $T_{\pm} = T \pm \tau_{\rm BR}/2$.   The total time-dependent Hamiltonian of the atoms is
\begin{equation}   \label{eq:H0+H_BR}
  H (t) = H_0 + H_{\rm Stark, BR} \, \Theta \Big( \frac{\tau_{\rm BR}}{2} - | t - T| \Big) .
\end{equation}

For $\hbar \Omega_{\rm BR} \ll {\mathcal E}_{0}$, the solution of the Schr\"odinger equation $i \hbar \frac{\partial \Psi ({\bf r}, t)}{\partial t} = H(t) \Psi ({\bf r}, t)$ is given by
\begin{eqnarray}
  \Psi ({\bf r}, t) &=&
  \sum_{n = -1, 0}
  {\mathcal X}_{n} (t) \, \psi_{{\bf k}_{n}} ({\bf r}) ,
\end{eqnarray}
where ${\mathcal X}_{0} (t) = \cos( \tfrac{\Omega_{\rm BR} t'}{2}) {\mathcal X}_{0} (T_-)
- i \, \sin( \tfrac{\Omega_{\rm BR} t'}{2}) {\mathcal X}_{-1} (T_-)$
and ${\mathcal X}_{-1} (t) = \cos( \tfrac{\Omega_{\rm BR} t'}{2}) {\mathcal X}_{-1} (T_-)
- i \, \sin( \tfrac{\Omega_{\rm BR} t'}{2}) {\mathcal X}_{0} (T_-)$, where $t' = t - T_-$.  The duration of the BR pulse is $\tau_{\rm BR} = \pi / \Omega_{\rm BR}$, and the spin wave functions ${\mathcal X}_{n} (T_{+})$ at $t = T_{+}$ are
\begin{equation}
  \left( \! \!
    \begin{array}{c}
      {\mathcal X}_{0} (T_{+}) \\ {\mathcal X}_{- 1} (T_{+})
    \end{array}
  \! \!
  \right) =
  \left( \! \!
    \begin{array}{cc}
      0 & - i \, {\mathbb I}
      \\
      - i \, {\mathbb I} & 0
    \end{array}
  \! \!
  \right) \,
  \left( \! \!
    \begin{array}{c}
      {\mathcal X}_{0} (T_{-}) \\ {\mathcal X}_{- 1} (T_{-})
    \end{array}
  \! \!
  \right) .
\end{equation}
The BR pulse reflects the atom beam with wave vector ${\bf k}_n = (k_x, (2n+1) K_0, 0)$, and the reflected beam has wave vector ${\bf k}_{-1-n} = (k_x, -(2n+1) K_0, 0)$, where $n = 0, -1$.

{\it Application of Mach-Zehnder interferometer to measure electric field strengths}---During the time interval $\frac{1}{2} \tau_{\rm BS} < t < T_-$, the atom propagates in a static homogeneous electric field ${\bf E}$.  The wavepacket propagating along paths $a$ and $b$ have wave vectors ${\bf k}_{0}$ and ${\bf k}_{-1}$ respectively.  The Hamiltonian for atoms propagating along paths $\nu = a, b$ 
\begin{equation}   \label{eq:H_nu}
  H = \frac{1}{2 M} \, \Big( {\bf p}+ \frac{g_F \mu_B}{4 c} \, {\bf F} \times {\bf E} \Big)^{2} ,
\end{equation}
where ${\bf p} = - i \hbar {\boldsymbol \nabla}$, the Land\'e $g$-factor is given by
$g_F = g_J \, \frac{F (F + 1) - I (I + 1) + J (J + 1)}{2 F (F + 1)} + 
g_I \, \frac{F (F + 1) + I (I + 1) - J (J + 1)}{2 F (F + 1)}$, $J = \frac{1}{2}$ is the total electronic angular momentum,
$I$ is the nuclear spin, $F = I \pm \frac{1}{2}$ is the total atomic angular momentum, $g_J$ is the electron $g$-factor, and $g_I$ is the nuclear $g$-factor.  It is convenient to introduce the unitary transformation of the Hamiltonian, $H = {\mathcal U}_{n_\nu} ( s({\bf r}) ) \tfrac{{\bf p}^2}{2M} {\mathcal U}^{-1}_{n_\nu} ( s({\bf r}) )$, with the unitary matrix
\begin{equation}   \label{eq:U_n-def}
  {\mathcal U}_{n_\nu} ( s({\bf r}) ) =
  \exp \Big( - \frac{i g_F \mu_B \, s ({\bf r})}{4 \hbar c} \, \big[ {\bf F} \times {\bf E} \big] \cdot \hat{\bf k}_{n_\nu} \Big) ,
\end{equation}
where $s({\bf r}) = \hat{\bf k}_{n_\nu} \cdot {\bf r} $ is the arc-length along path $\nu$, the integer number $n_\nu$ is defined as $n_a = 0$ and $n_b = -1$, and $\hat{\bf k}_{n_\nu} = {\bf k}_{n_\nu} / k_0$.

The atom wavepacket at $t = \tau_{\rm BS}$ is taken to be centered at the origin of the coordinate system, ${\bf R} (0) = {\bf 0}$.  At $t > \tau_{\rm BS}$, the wavepacket with wavevector ${\bf k}_{n}$ is centered at ${\bf R} (t) = {\bf v}_{n} t$, where ${\bf v}_{n} = \frac{\hbar}{M} {\bf k}_{n}$ is the group velocity.  The arc-length $s (t)$ can be defined as $s (t) = {\bf k}_{n} \cdot {\bf R} (t) = v_0 t$, where $v_0 = \frac{\hbar}{M} | {\bf k}_n|$.  The solution of the Schr\"odinger equation $i \hbar \frac{\partial}{\partial t} \Psi_{\nu} ({\bf r}, t) = H \Psi_{\nu} ({\bf r}, t)$ for $\frac{\tau_{\rm BS}}{2} < t < T_{-}$ can be written as
$\Psi_{\nu} ({\bf r}, t) = e^{i {\bf k}_{n_\nu} \cdot {\bf r} - i \epsilon_{k_0} t / \hbar}
{\mathcal X}_{\nu} (s ({\bf r}))$, where $\nu = a, b$ for beams $a$ and $b$,
${\mathcal X}_{\nu} (s ({\bf r})) = {\mathcal U}_{n_\nu} (s ({\bf r})) {\mathcal X}_{n_\nu} ( \frac{\tau_{\rm BS}}{2} )$,
${\mathcal X}_{0} (\frac{\tau_{\rm BS}}{2})$ and ${\mathcal X}_{-1} (\frac{\tau_{\rm BS}}{2})$ are given in Eq.~(\ref{spin-wave-function-BS1}).
Note that ${\bf k}_{n_\nu} \cdot {\bf r} = k_0 s ({\bf r})$, therefore $\Psi_{\nu} ({\bf r}, t)$ is a function of the arc-length $s ({\bf r})$ and time $t$.
At $t = T_-$, the spin wave function is
\begin{equation}   \label{eq:X_nu-T}
  {\mathcal X}_{\nu} (s_T) = {\mathcal U}_{n_\nu} (s_T) {\mathcal X}_{\nu} \Big( \frac{\tau_{\rm BS}}{2} \Big) ,
\end{equation}
where $s_T = \frac{\hbar}{M} k_0 T$ is path that the wave packet propagates over during the time $T$.

As described above where the BR was considered, within the time interval $T_{-} < t < T_{+}$, the BR pulse reflects the beam with the wave vector ${\bf k}_{n}$ (with $n = 0, -1$) such that the reflected beam propagates with the wave vector ${\bf k}_{-1-n}$.   Hereafter $\tau_{\rm BS}$ and $\tau_{\rm BR}$ are assumed to be small, therefore the phase shifts $\frac{\hbar}{M} k_0 \tau_{\rm BS} \ll 1$ and $\frac{\hbar}{M} k_0 \tau_{\rm BR} \ll 1$ can be neglected.

The BR pulse reflects the atom beams with wave vectors ${\bf k}_{n}$ with $n = 0, -1$, and the reflected beam has wave vector ${\bf k}_{-1-n}$.
The wave functions $\Psi_{\nu} ({\bf r}, t)$ for $T_{+} < t < 2 T - \frac{\tau_{\rm BS}}{2}$ is
$\Psi_{\nu} ({\bf r}, t) = - \frac{i}{\sqrt{2}} e^{i k_0 {\tilde s}({\bf r}) - i \epsilon_{k_0} t / \hbar} {\mathcal X}_{\nu} ({\tilde s}({\bf r}))$,
where
${\mathcal X}_{\nu} ({\tilde s}({\bf r})) = {\mathcal U}_{-1-n_\nu} ({\tilde s}({\bf r}) - s_T) {\mathcal X}_{\nu} ( T_{-} )$, the arc-length is ${\tilde s}({\bf r}) = s_T + s ({\bf r} - {\bf v}_{n} T)$, the path $s_T$ the wave propagates at time $T$ is $s_T = | {\bf v}_n | T$, and the group velocity is ${\bf v}_{n} = \frac{\hbar}{M} {\bf k}_{n}$.  The path $s_{2T}$ the wave propagates at time $2T$ is $s_{2T} = 2 |{\bf v}_{n}| T$.
Using Eq.~(\ref{eq:X_nu-T}), ${\mathcal X}_{\nu} (s_{2T})$ with ${\tilde s}({\bf r}) = s_{2T}$ can be rewritten as
\begin{equation}   \label{eq:spin-wave-function-t>T-total}
  {\mathcal X}_{\nu} (s_{2T}) =
  {\mathcal U}_{-1-n_\nu} (s_T) \,
  {\mathcal U}_{n_\nu} (s_T) \,
  {\mathcal X}_{\nu} \Big( \frac{\tau_{\rm BS}}{2} \Big) .
\end{equation}

The second BS is switched on at $t = 2 T - \tau_{\rm BS}/2$ and switched off at $t = 2 T + \tau_{\rm BS}/2$.
The wave functions $\Psi_{\mu} ({\bf r}, 2 T + \tau_{\rm BS}/2)$ propagating along paths $\mu = c, d$ (see Fig.~\ref{Fig:Bragg-splitter}) are
\begin{equation}   \label{eq:Psi_cd-2T}
  \Psi_{\mu} \Big( {\bf r}, 2 T + \frac{1}{2} \tau_{\rm BS} \Big) =
  {\mathcal X}_{\mu} \,
  \psi_{{\bf k}_{n_\mu}} ({\bf r}) ,
\end{equation}
where ${\mathcal X}_{\mu} \equiv {\mathcal X}_{n_{\mu}} ( 2 T + \frac{\tau_{\rm BS}}{2} )$, and
$n_{\mu}$ is defined as: $n_c = 0$ and $n_d = -1$.  The spin wave functions ${\mathcal X}_\mu$ are found from Eq.~(\ref{eq:for-X-BS-2-levels}),
\begin{subequations}   \label{subeqs:X-BS2}
\begin{eqnarray}
  \label{eq:Xc-BS2}
  {\mathcal X}_{c} &=&
  {\mathcal V}_{T} \, {\mathcal X}_{0} ( s_{2 T} ) +
  {\mathcal V}_{R} \, {\mathcal X}_{-1} ( s_{2 T} ) ,
  \\
  \label{eq:Xd-BS2}
  {\mathcal X}_{d} &=&
  - {\mathcal V}_{R} \, {\mathcal X}_{0} ( s_{2 T} ) +
  {\mathcal V}_{T} \, {\mathcal X}_{-1} ( s_{2 T} ) ,
\end{eqnarray}
\end{subequations}
where ${\mathcal V}_{T}$ and ${\mathcal V}_{R}$ are given in Eq.~(\ref{eq:U_BS-V_BS}).  Note that the second BS splits each of the wavepackets propagating along paths $a$ and $b$ into the wavepackets propagating along paths $c$ and $d$, see Fig.~\ref{Fig:Bragg-splitter}.  Let us consider the wavepacket in Eq.~(\ref{eq:Xc-BS2}) which propagates along path $c$. The first term on the right-hand side of Eq.~(\ref{eq:Xc-BS2}) describes the wavepacket that originates from path $b$ and has wavevector ${\bf k}_{0}$.  The second term is the wavepacket from path $a$ and has wavevector ${\bf k}_{-1}$.  Similarly, the first and second terms on the right-hand side of Eq.~(\ref{eq:Xd-BS2}) are wavepackets propagating along path $d$ that originate from paths $b$ and $a$.  Using Eqs.~(\ref{spin-wave-function-BS1}) and (\ref{eq:spin-wave-function-t>T-total}), Eq.~(\ref{subeqs:X-BS2}) become
\begin{equation}  \label{eq:X_cd}
  {\mathcal X}_{c} =
  {\mathcal W}_{ac}  \, \chi_1 +
  {\mathcal W}_{bc} \, \chi_1 ,
  \quad
  {\mathcal X}_{d} =
  {\mathcal W}_{ad}  \, \chi_1 +
  {\mathcal W}_{bd}  \, \chi_1 ,
\end{equation}
where ${\mathcal W}_{ac} = - i \, {\mathcal V}_{R} \, {\mathcal U}_{-1} \, {\mathcal U}_{0} \, {\mathcal V}_{T}$,
${\mathcal W}_{bc} = i \, {\mathcal V}_{T} \, {\mathcal U}_{0} \, {\mathcal U}_{-1} \, {\mathcal V}_{R}$,
${\mathcal W}_{ad} = - i \, {\mathcal V}_{T} \, {\mathcal U}_{-1} \, {\mathcal U}_{0} \, {\mathcal V}_{T}$,
${\mathcal W}_{bd} = - i \, {\mathcal V}_{R} \, {\mathcal U}_{0} \, {\mathcal U}_{-1} \, {\mathcal V}_{R}$,
and ${\mathcal U}_{n} \equiv {\mathcal U}_{n} (s_T)$ for $n = 0, -1$.

In the absence of an electric field, ${\mathcal U}_{n} (s_T) = {\mathbb I}$ [see Eq.~(\ref{eq:U_n-def})], so ${\mathcal X}_{c} = 0$, ${\mathcal X}_{d} = - i \chi_x$, and $| {\mathcal X}_{d} |^{2} = 1$.  In the presence of the electric field, $| {\mathcal X}_{d} |^{2}$ can be written as
\begin{eqnarray}
  | {\mathcal X}_{d} |^{2} &=&
  \chi_1^{\dag}
  \big(
    {\mathcal W}_{ad}^{\dag} \, {\mathcal W}_{ad} +
    {\mathcal W}_{bd}^{\dag} \, {\mathcal W}_{bd}
    \nonumber \\ && +
    {\mathcal W}_{ad}^{\dag} \, {\mathcal W}_{bd} +
    {\mathcal W}_{bd}^{\dag} \, {\mathcal W}_{ad}
  \big)
  \chi_1 .
\end{eqnarray}
The real terms $\chi_z^{\dag} {\mathcal W}_{ad}^{\dag} {\mathcal W}_{ad} \chi_z$ and
$\chi_z^{\dag} {\mathcal W}_{bd}^{\dag} \, {\mathcal W}_{bd} \chi_z$ are probabilities to find the atom on path $d$ after passing path $a$ and $b$, respectively.  The complex terms $\chi_z^{\dag} {\mathcal W}_{ad}^{\dag} {\mathcal W}_{bd} \chi_z$ and $\chi_z^{\dag} {\mathcal W}_{bd}^{\dag} \, {\mathcal W}_{ad} \chi_z$, which are complex conjugates, are interference terms which contain the phase difference of the wavepackets, $\varphi_{\rm AC}$, after propagating on paths $a$ and $b$.  This phase difference is the AC phase.

Writing $\chi_1^{\dag} {\mathcal W}_{ad}^{\dag} {\mathcal W}_{bd} \chi_1 = | \chi_1^{\dag} {\mathcal W}_{ad}^{\dag} {\mathcal W}_{bd} \chi_1 | \, e^{i \varphi_{\rm AC}}$, the AC phase can be computed as $\varphi_{\rm AC} = {\rm Arg} [ \chi_1^{\dag} {\mathcal W}_{ad}^{\dag} {\mathcal W}_{bd} \chi_1 ]$.
For $|{\bf E}| \ll | \tfrac{4 \hbar c}{g_F \mu_B s_T} |$,
\begin{equation}   \label{eq:AC_phase}
  \varphi_{\rm AC} =
  \frac{g_F \mu_B s_T}{\sqrt{2} \, \hbar c} \, \frac{k_x}{\sqrt{k_x^2 + K_0^2}} \, E_y ,
\end{equation}
see ``Computation of the AC phase'' in the End Matter.
Note that measuring the AC phase $\varphi_{\rm AC}$ and applying Eq.~(\ref{eq:AC_phase}), one can compute the electric field strength $E_y$.


It is important to note that with spin-independent BSs, the AC phase is quadratic in $E$, whereas, with SDO BSs, the AC phase is linear in $E$. Consequently, the accuracy and precision of measurement of the electric field strength is vastly improved.

{\it Summary and Conclusions}---Atom (and molecule) BSs create superpositions of momentum states and control the flow of the paths of atomic beams.  These devices are indispensable for exploring fundamental quantum phenomena, developing new quantum technologies.  They are crucial components in atom interferometry, enabling precise measurements of fundamental constants, gravity, and rotations.   SDO BSs also create superpositions of internal atomic states in their output beams.  Here, an application for SDO lattices that uses them as a SDO BS has been proposed. SDO BSs create a superposition of internal spin states as well as momentum states.  We have discussed one utilization of a SDO BS, namely, the measurement of external electric field strengths.



\section{End Matter}

\subsection{Scalar and vector polarizabilities}
The scalar polarizability $\alpha_s (\omega)$ and the vector polarizability $\alpha_v (\omega)$ are given by~\cite{SOI-EuroPhysJ-13, comment-tensor},
\begin{eqnarray}
  \label{eq:alpha_s}
  \alpha_s (\omega) &=&
  \frac{1}{\sqrt{3 (2 J + 1)}} \, \alpha_0 (\omega) ,
  \\
  \label{eq:alpha_v}
  \alpha_v (\omega) &=&
  \big( - 1 \big)^{J + I + F} \,
  \sqrt{\frac{2 F (2 F + 1)}{F + 1}}
  \nonumber \\ && \times
  \left\{
    \begin{array}{ccc}
      F & 1 & F
      \\
      J & I & J
    \end{array}
  \right\} \,
  \alpha_1 (\omega) ,
\end{eqnarray}
where $\left\{  \begin{array}{ccc} F & 1 & F \\ J & I & J \end{array} \right\}$ is a Wigner 6-$j$ symbol.  The reduced dynamical polarizabilities are~\cite{SOI-EuroPhysJ-13},
\begin{eqnarray}   \label{eq:alpha_K}
  \alpha_K (\omega) &=&
  \frac{\sqrt{2 K + 1}}{\hbar} \,
  \sum_{n' J'}
  \big( - 1 \big)^{K + J' + 3/2}
  \left\{
    \begin{array}{ccc}
      1 & K & 1
      \\
      \frac{1}{2} & J' & \frac{1}{2}
    \end{array}
  \right\}
  \nonumber \\ && \times
  \big| \langle n' 1 J' \| -e \, r \| n 0 \tfrac{1}{2} \rangle \big|^{2}
  \nonumber \\ && \times
  {\rm Re}
  \bigg\{
    \frac{1}{\omega_{n' 1 J', n 0 \frac{1}{2}} - \omega - \frac{i}{2} \gamma_{n' 1 J'}}
    \nonumber \\ && +
    \frac{( - 1)^{K}}{\omega_{n' 1 J', n 0 \tfrac{1}{2}} + \omega + \frac{i}{2} \gamma_{n' 1 J'}}
  \bigg\} ,
\end{eqnarray}
where $K = 0$ for the reduced dynamical scalar polarizability and $K = 1$ for the reduced dynamical vector polarizability, $|n 0 \frac{1}{2}\rangle$ is the wave function of the alkali atom ground $s$ state, $|n' 1 J' \rangle$ is the wave function of an excited  $p$ state, the transition frequency is represented by $\omega_{n' 1 J', n 0 \frac{1}{2}}$, the line widths are represented $\gamma_{n' 1 J'}$, and $\langle n' L' J' \| -e \,  r \| n L J \rangle$ is a reduced matrix element of the electric dipole moment operator.

\subsection{Wave function for BS Hamiltonian in Eq.~(\ref{eq:Schrodinger-BS})}

The wave function $\Psi ({\bf r}, t)$ in the Schr\"odinger equation (\ref{eq:Schrodinger-BS}) can be expanded as
\begin{equation}   \label{eq:wave-function-general}
  \Psi ({\bf r}, t) =  e^{- i \epsilon_{k_0} t / \hbar}
  \sum_{n = - \infty}^{\infty}
  {\mathcal X}_{n} (t) \, \psi_{{\bf k}_{n}} ({\bf r}) ,
\end{equation}
where the spin wave functions ${\mathcal X}_{n} (t)$ resulting from the Bragg scattering are determined from the differential equations
\begin{eqnarray}   \label{eq:for-A}
  i \, \frac{d {\mathcal X}_{n} (t)}{d t} &=&
  \frac{1}{\hbar} \, \big( \epsilon_{k_n} - \epsilon_{k_0} \big) {\mathcal X}_{n} (t) -
  i \, \Omega_{\rm BS}
  F_y
  \nonumber \\ && \times
  \big[ {\mathcal X}_{n + 1} (t) - {\mathcal X}_{n - 1} (t) \big] ,
\end{eqnarray}
with the amplitudes satisfying the initial conditions
\begin{equation}   \label{eq:A_ini}
  {\mathcal X}_{n} \Big( - \frac{\tau_{\rm BS}}{2} \Big) =
  \delta_{n, 0} \, \chi_1 .
\end{equation}

If $\hbar \Omega_{\rm BS} \ll {\mathcal E}_{0}$, the probability of quantum transition to the high-energy states with $n \geq 1$ or $n \leq - 2$ can be neglected, and the chain of equations (\ref{eq:for-A}) reduces to Eq.~(\ref{eq:for-X-BS-2-levels}) in the main text.

\subsection{Computation of the AC phase}

Using equations for ${\mathcal W}_{ad}$ and ${\mathcal W}_{bd}$ after Eq.~(\ref{eq:X_cd}), the matrix ${\mathcal W}_{a d}^{\dag} \, {\mathcal W}_{b d}$ can be written as
\begin{equation}   \label{eq:W_ad^dag-W_bd}
  {\mathcal W}_{a d}^{\dag} \, {\mathcal W}_{b d} =
  {\mathcal V}_{T} \, {\mathcal U}_{0}^{\dag} \, {\mathcal U}_{-1}^{\dag} \, {\mathcal V}_{T} \,
  {\mathcal V}_{R} \, {\mathcal U}_{0} \, {\mathcal U}_{-1} \, {\mathcal V}_{R} ,
\end{equation}
where ${\mathcal V}_{T}$ and ${\mathcal V}_{R}$ are given in Eq.~(\ref{eq:U_BS-V_BS}), and the unitary matrices ${\mathcal U}_{n}$ with $n = 0, -1$ are given in Eq.~(\ref{eq:U_n-def}) with $s ({\bf r}) = s_T$.
For weak electric field,
\begin{equation}   \label{eq:U_n-series}
  {\mathcal U}_{n} \approx
  {\mathbb I} -
  \frac{i g_F \mu_B \, s_T}{4 \hbar c} \, \big[ {\bf F} \times {\bf E} \big] \cdot \hat{\bf k}_{n} .
\end{equation}
Substituting this equation into Eq.~(\ref{eq:W_ad^dag-W_bd}) and neglecting quadratic terms with the electric field results in
\begin{eqnarray}   \label{eq:W_ad^dag-W_bd-series}
  {\mathcal W}_{a d}^{\dag} \, {\mathcal W}_{b d} &\approx&
  {\mathcal V}_{T}^{2} \, {\mathcal V}_{R}^{2} -
  \frac{i g_F \mu_B \, s_T}{4 \hbar c}
  \nonumber \\ &\times&
  \big[ \boldsymbol{\mathcal F} \times {\bf E} \big] \cdot \big( \hat{\bf k}_{0} + \hat{\bf k}_{-1} \big) ,
\end{eqnarray}
where
\begin{equation}   \label{eq:VT2VRFVR-VTFVTVR2-def}
  \boldsymbol{\mathcal F} =
  {\mathcal V}_{T}^{2} \, {\mathcal V}_{R} \, {\bf F} \, {\mathcal V}_{R} -
  {\mathcal V}_{T} \, {\bf F} \, {\mathcal V}_{T} \, {\mathcal V}_{R}^{2} .
\end{equation}
Using Eq.~(\ref{eq:U_BS-V_BS}),  the following equations can be obtained,
\begin{eqnarray}
  &&
  {\mathcal V}_{T}^{2} \, {\mathcal V}_{R}^{2} =
  \frac{1}{6} \, {\mathbb I} + \frac{1}{8} \, Q_{y, y} ,
  \quad
  {\mathcal V}_{T}^{2} \, {\mathcal V}_{R} \, {\bf F} \, {\mathcal V}_{R} =
  \frac{1}{4} \, F_y \, \hat{\bf y} ,
  \\
  &&
  {\mathcal V}_{T} \, {\bf F} \, {\mathcal V}_{T} \, {\mathcal V}_{R}^{2} =
  \frac{i}{4 \sqrt{2}} \, Q_{y z} \, \hat{\bf x} +
  \frac{1}{4} \, F_y \, \hat{\bf y}
  \nonumber \\ && \qquad \qquad \qquad +
  \frac{1}{4 \sqrt{2}} \, \big( F_z - i \, Q_{x, y} \big) \, \hat{\bf z} .
\end{eqnarray}
Then Eq.~(\ref{eq:VT2VRFVR-VTFVTVR2-def}) takes the form
\begin{eqnarray}   \label{eq:VT2VRFVR-VTFVTVR2-res}
  \boldsymbol{\mathcal F} &=&
  - \frac{1}{4 \sqrt{2}} \,
  \Big[
    i \, Q_{y z} \, \hat{\bf x} +
    \big( F_z - i \, Q_{x y} \big) \, \hat{\bf z}
  \Big] .
\end{eqnarray}
The unit vectors $\hat{\bf k}_{0}$ and $\hat{\bf k}_{-1}$ are given after Eq.~(\ref{eq:U_n-def}), and their sum is
\begin{equation}   \label{eq:hat-k_0+hat-k_1}
  \hat{\bf k}_{0} + \hat{\bf k}_{-1} = \frac{2 k_x}{\sqrt{k_x^2 + K_0^2}} \, \hat{\bf x} .
\end{equation}
Substituting Eqs.~(\ref{eq:VT2VRFVR-VTFVTVR2-res}) and (\ref{eq:hat-k_0+hat-k_1}) into Eq.~(\ref{eq:W_ad^dag-W_bd-series}), the following equation for ${\mathcal W}_{a d}^{\dag} \, {\mathcal W}_{b d}$ is obtained
\begin{eqnarray}   \label{eq:W_ad^dag-W_bd-res}
  {\mathcal W}_{a d}^{\dag} \, {\mathcal W}_{b d} &\approx&
  {\mathcal V}_{T}^{2} \, {\mathcal V}_{R}^{2} +
  \frac{i g_F \mu_B \, s_T}{8 \sqrt{2} \, \hbar c} \,
  \frac{k_x}{\sqrt{k_x^2 + K_0^2}}
  \nonumber \\ &\times&
  \big( F_z - i \, Q_{x y} \big) \, E_y .
\end{eqnarray}
The matrix element $\chi_1^{\dag} {\mathcal W}_{a d}^{\dag} \, {\mathcal W}_{b d} \chi_1$ is
\begin{equation}   \label{eq:WW-matrix-element}
  \chi_1^{\dag} {\mathcal W}_{a d}^{\dag} \, {\mathcal W}_{b d} \chi_1 =
  \frac{1}{8} +
  \frac{i g_F \mu_B \, s_T}{8 \sqrt{2} \, \hbar c} \,
  \frac{k_x}{\sqrt{k_x^2 + K_0^2}} \,
  E_y .
\end{equation}
The application of the  equations prior to Eq.~(\ref{eq:AC_phase}) into Eq.~(\ref{eq:WW-matrix-element}), results in Eq.~(\ref{eq:AC_phase}) in the main text.
If initially the spin wave function is $\chi_m$ with $m = 0, \pm 1$, the matrix element of ${\mathcal W}_{a d}^{\dag} \, {\mathcal W}_{b d}$ is
\begin{equation*}
  \chi_m^{\dag} {\mathcal W}_{a d}^{\dag} \, {\mathcal W}_{b d} \chi_m \approx
  \frac{2 - m^2}{8} +
  \frac{i \, m}{8} \, \varphi_{\rm AC} ,
\end{equation*}
therefore the AC phase is $\varphi_m \approx m \, \varphi_{\rm AC}$. If $m = 0$, the AC phase is quadratic with electric field strength.


\begin{thebibliography}{99}
\bibitem{comment-tensor}
If an atom in the ground state has $L \neq 0$, the effective interaction of an atom with the electromagnetic field consists of a scalar potential, a vector potential containing what Cohen-Tannoudji and Dupont-Roc termed the fictitious magnetic field, and a tensor potential.  For alkali atoms in the ground state, $J = \frac{1}{2}$, and the tensor potential vanishes.
\bibitem{comment:omega_1}
The frequency $\omega_1$ is such that $\alpha_s (\omega_1) = | \alpha_v (\omega_0) |$, hence the Rabi frequencies, $\Omega_{\rm BS}$ and $\Omega_{\rm BR}$, given in Eqs.~(\ref{eq:Rabi-frequency_v}) and (\ref{eq:Rabi-frequency_s}), are equal to each other.
\end{thebibliography}
\end{document}